# On the general structure of mathematical models for physical systems

D. H. Delphenich [†]
2801 Santa Rosa Dr., Kettering, OH 45440 USA

It is proposed that the mathematical models for any physical systems that are based in first principles, such as conservation laws or balance principles, have some common elements, namely, a space of kinematical states, a space of dynamical states, a constitutive law that associates dynamical states with kinematical states, as well as a duality principle.  The equations of motion or statics then come about from, on the one hand, specifying the integrability of the kinematical state, and on the other hand, specifying a statement that is dual to it for the dynamical states. Examples are given from various fundamental physical systems.

## Contents

## 1  Introduction

When one surveys the various mathematical models for physical systems, and especially the ones that are based in "first principles," one eventually sees that a common pattern emerges.  One can characterize the basic elements of a mathematical model for a physical system – whether static or dynamic – as generally involving five basic components:

---

[†]  E-mail: david_delphenich@yahoo.com

1. A space of kinematical states, including the constraints imposed on it.
2. An equation or system of equations that specify the "degree of integrability" of the kinematical states.
3. A space of dynamical states that is, in some definable sense, "dual" to the space of kinematical states.
4. A constitutive law that associates a dynamical state with a kinematical state.
5. An equation or system of equations that describes the conservation law or balance principle that the dynamical state is subject to.

The main purpose of this study is to show that the geometry of jet manifolds presents a natural methodology for accomplishing all of those objectives in a single formalism that also borders on other branches of mathematics that are important to the problem of modeling physical systems, such a the calculus of variations and the theory of differential equations, both ordinary and partial. After presenting the general mathematical theory, we shall attempt to justify this claim by exhibiting how it would apply to various physical models of varying degrees of complexity. However, we shall also show how the same basic elements can be applied in the more familiar cases in which the integrability relates to the exterior derivative operator.

Of particular significance is the fact that when one uses the "Spencer operator" that is associated with the integrability of sections of the projection of a jet manifold onto its source manifold as a way of defining the degree of integrability of kinematical states − viz., sections of that type − one finds that the most natural definition of associated dynamical states, namely, 1-forms of a certain type on the jet manifold, includes not only the introduction of a set of constitutive laws, but also makes the conservation law or balance principle that the dynamical state obeys take the form of an "adjoint" to the integrability law for the kinematical states. The duality under which the adjoint is defined is basically the concept of virtual work. Thus, the real innovation in the model is that it relates the first principles of physics that take the form of balance or conservation laws to something purely mathematical, such as the integrability of kinematical states. One then finds that the general technique applies to physical systems that are concerned with the statics or dynamics of points, extended matter distributions, or fields. Furthermore, unlike the various models that start with energy as their fundamental concept, virtual work is just as useful when dealing with things that lie beyond the scope of energy methods, such as non-conservative forces and non-holonomic constraints. One also starts to see how the integrability of kinematical states might also have a subtle relationship with the issue of whether the physical system is "open" or "closed;" i.e., the non-integrability might be related to the error in the approximation that made the system model closed.

In the first section of this paper, we summarize the geometry of jet manifolds in a form that is closest to the spirit of most physical models, which is to say, systems of differential equations. In the next section, we define the general model for physical systems that comes out of that methodology. In the remaining section, we exhibit how the general model specializes to various physical systems, such as the motion of points, extended matter distributions, waves, and electromagnetic fields. In the case of electromagnetism, we shall show that the same basic structural features can be attributed to the integrability of the exterior derivative operator. We then conclude with a discussion of how one might further explore the application of the more exotic mathematical techniques that grow out of the present methodology in a manner that seems rooted in physical considerations.

Many of the basic ideas that are discussed in what follows have been treated by the author in previous papers [**1-5**] in various forms. The advance of the present paper over them is to collect those individual results into a general methodology.

## 2 The geometry of jets

The roots of the geometry of jets are in contact geometry, which generalizes the basic geometric notion of tangency. One says that two curves, for instance, have *zero-order contact* at a point if they intersect, *first-order contact* if they intersect and share a common tangent, *second-order contact* if the intersect, share a common tangent, and also share a common "osculating conic," and so on. What one is basically doing is to go to successively higher derivatives of the position of the point in space with respect to the curve parameter in order to define the order of contact between the curves. Similarly, when one is concerned with higher-dimensional objects than points that require more than one parameter to define, one can use successive partial derivatives.

One encounters the first step in the sequence of definitions of higher-order jets in the elementary theory of differentiable manifolds when one defines a tangent vector at a point to be an equivalence class of differentiable curves through that point that share a common tangent vector in $\mathbf{R}^n$ to that point for some, and thus any, coordinate chart about that point, where tangency in $\mathbf{R}^n$ is defined in the usual way that one learns in multi-variable calculus. However, nowadays, geometry generally does not go on to define equivalence classes of curves through the point that also share a common conic at that point, along with the tangent, or perhaps equivalence classes of surfaces through that point that share a common tangent plane, etc., where the next order of contact is an "osculating quadric." Rather, one simply switches to the analytical description of the equivalence classes, which is by means of the values and successive derivatives of the curves, surfaces, etc. through the point in question, as described in some, and thus any, coordinate system.

### 2.1 Jet manifolds

The notion of a "jet" was introduced by Ehresmann [**6**, **7**] in 1951 as a way of generalizing the differential-geometric notions of "infinitesimal structures" and "geometrical objects." It was then developed and applied to not only differential geometry in mathematics, but also differential topology, differential equations, and the calculus of variations. For a more modern treatment of the geometry of jet manifolds, one might confer Saunders [**8**].

To begin with, if $f: U \subset M \to N$, $u \mapsto x(u)$ is a continuous function at a point $u \in U$ then the *germ* of $f$ at $u$ is the set $j^0 f_u$ of all continuous functions $f': U' \subset M \to N$, such that $U'$ is also a neighborhood of $x$ and $f'(u) = f(u)$. Note that is it superfluous to specify the behavior of these functions globally on $M$ since the definition is purely local. Thus, germs can exist even when global extensions of the functions in question do not exist, such as one finds with local frame fields on non-parallelizable manifolds. One sees that the germ $j^0 f_u$ is completely characterized by the ordered pair of points $(u, x) \in M \times N$, which one might also denote by $J^0(M, N)$.

The next order of contact is the 1-jet of $f$ at $u$: If one now assumes that $f$ is continuously differentiable (i.e., $C^1$) at $x$ then the *1-jet* of $f$ at $u$ is the set $j^1 f_u$ that consists of all $C^1$ functions in $j^0 f_u$ that also satisfy $df|_u = df\lceil_u$. Generally, $j^1 f_u$ will be a proper subset of $j^0 f_u$ since one usually has functions that are continuous, but not differentiable, in a neighborhood of $u$. One can then uniquely characterize $j^1 f_u$ by the ordered triple of points $(u, x, df|_u) \in M \times N \times \mathrm{Hom}(T_u M, T_x N)$, in which $\mathrm{Hom}(T_u M, T_x N)$ is the vector space that is composed of all linear maps from the tangent space $T_u M$ to the tangent space $T_x N$. Thus, if $M$ is $m$-dimensional and $N$ is $n$-dimensional $\mathrm{Hom}(T_u M, T_x N)$ is $mn$-dimensional, while $M \times N \times \mathrm{Hom}(T_u M, T_x N)$ has dimension $m + n + mn$.

If one introduces a coordinate chart $(U, u^a)$ about $u \in M$ and another $(V, x^i)$ about $x \in N$ then the function $f: U \to V$ can be described by a system of $n$ equations in $m$ independent variables that takes the form:

$$x^i = x^i(u^a). \tag{2.1}$$

Although the partial derivatives of $x^i$ with respect to $u^a$ define a matrix of *functions* $x^i_{,a}(u) = \partial x^i / \partial u^a (u)$ on $U$, nonetheless, at the point $u$ they define only a matrix of *scalars* that we shall denote by $x^i_a$, in general. One then forms the coordinate representation of the 1-jet $(u, x, df|_u)$ at $u$ as:

$$j^1 x_u = (u^a, x^i, x^i_a). \tag{2.2}$$

If one defines $J^1(M, N)$ to be the set of all 1-jets of $C^1$ functions from $M$ to $N$ then one sees that choosing a coordinate chart $(U, u^a)$ about $u$ and another one $(V, x^i)$ about $x$, as above, allows one to also define matrix representations $x^i_a$ for the differential maps from each $u \in U$ to each $x \in V$ and one can regarded the functions $(u^a, x^i, x^i_a) : U \times V \times \mathrm{Hom}(T_u M, T_x N) \subset J^1(M, N) \to \mathbf{R}^m \times \mathbf{R}^n \times \mathbf{R}^{mn}$ as a coordinate chart on $J^1(M, N)$ about $j^1 x_u$. Thus, the set $J^1(M, N)$ becomes a differentiable manifold of dimension $m + n + mn$ that one calls the *manifold of 1-jets of $C^1$ maps from $M$ to $N$*. Generally, if $(\bar{u}^a, \bar{x}^i, \bar{x}^a_i)$: $U' \times V' \times \mathrm{Hom}(T_u M, T_x N) \subset J^1(M, N) \to \mathbf{R}^m \times \mathbf{R}^n \times \mathbf{R}^{mn}$ is another coordinate chart about $u$ then the transformation of coordinates on the intersection of the charts comes about by a transformation of the form:

$$\bar{u}^a = \bar{u}^a(u^b), \quad \bar{x}^i = \bar{x}^i(x^j), \quad \bar{x}^i_a(\bar{u}, \bar{x}) = \bar{x}^i_{,j}(x) x^j_b(u, x) \bar{u}^b_{,a}(u). \tag{2.3}$$

Note that more general transformations of $\bar{x}^i_a(\bar{u}, \bar{x})$ are possible if one replaces the invertible matrices $\bar{x}^i_{,j}(x)$ and $\bar{u}^b_{,a}(u)$ with invertible matrices $\bar{x}^i_j(x)$ and $\bar{u}^b_a(u)$ that are not required to be the Jacobian matrices of the coordinate transformations.

This brings us close to the recurring issue of integrability, which requires that we first define the three basic projections that are defined by any jet manifold, such as $J^1(M, N)$:

1. The *source* projection:     $\alpha: J^1(M, N) \to M$,     $j^1 x_u \mapsto u$,
2. The *target* projection:     $\beta: J^1(M, N) \to N$,     $j^1 x_u \mapsto x$,
3. The *contact* projection:     $\pi_0^1: J^1(M, N) \to M \times N$,     $j^1 x_u \mapsto (u, x)$.

In general, the first two projections do not define fibrations, since the transformation of local sections does behave properly for a local trivialization, and one thus thinks of them as simply *fibered manifolds*. The contact projection defines an affine bundle over $M \times N$, whose fibers are $\text{Hom}(T_u M, T_x N)$, which are thus modeled on the vector space $\text{Hom}(\mathbf{R}^m, \mathbf{R}^n)$; the reason that one says "affine," instead of "vector" relates to the nature of the local transformation laws for the local sections, which involve the addition of a term, along with matrix multiplications.

The reason that we are calling the third projection the "contact" projection is because that would make sections of that projection take the form of fields of contact elements. That is, since the differential map $df|_u : T_u M \to T_x N$ is a linear map, the image of $T_u M$ will be a linear subspace of $T_x N$ whose dimensions does not exceed $m$. In general, we shall refer to any linear subspace of a tangent space as a *contact element* at that point. The relationship to the matrices of $\text{Hom}(\mathbf{R}^m, \mathbf{R}^n)$ is that if $x_a^i$ is such a matrix and one has coordinates $u^a$ and $x^i$ about the points $u$ and $x$, as above, then the linear map $T : T_u M \to T_x N$ that is defined by:

$$T = x_a^i \, du^a \otimes \frac{\partial}{\partial x^i} \tag{2.4}$$

also defines a contact element at $x$ whose dimension does not exceed $m$.

One immediately sees that the manifold $J^1(M, N)$ generally contains contact elements of varying dimensions that range from 0 to $m$. If one wishes to specify that only contact elements of dimension $m$ must appear then not only must $m$ be no greater than $n$, but one must consider only 1-jets of local *immersions*, whose differential maps will have the maximum rank. Once again, it is superfluous to strengthen this to local embeddings, since the difference between immersions and embeddings is of a global nature. Ultimately, when $m \leq n$ if one wishes to consider only contact elements in $N$ of dimension $m$ then one must restrict oneself to a submanifold of $J^1(M, N)$ that is defined by the level set of $m$ under the integer function that takes any $j^1 x_u$ to the rank of the linear map from $T_u M$ to $T_x N$ that is associated with that jet.

Before we go on to introducing more advanced notions, we shall first give some common examples of jet manifolds that occur in physics.

The manifold $J^1(\mathbf{R}, M)$ consists of 1-jets of continuously differentiable curves in the manifold $M$. Its coordinate charts look like $(t, x^i, v^i)$, where the numbers $v^i$ are a generalization of the components of the velocity vector to some curve $x(t)$ through point $x$ at the time $t$, except that velocity would represent the "integrable" possibility, which we shall discuss shortly. One also sees that $J^1(\mathbf{R}, M)$ also locally looks like $\mathbf{R} \times T(M)$.

The manifold $J^1(M, \mathbf{R})$ consists of 1-jets of continuously differentiable functions on the manifold $M$. Its coordinate charts look like $(x^i, \phi, \phi_i)$, in which the coordinates $\phi_i$ similarly generalize the components of the 1-form $d\phi$, with a similar comment regarding integrability. Dual to the last example, the manifold $J^1(M, \mathbf{R})$ locally looks like $T^*M \times \mathbf{R}$.

The manifold $J^1(\mathbf{R}^k, M)$, when restricted to the immersions, defines *k-frames* in the tangent spaces to $M$; the contact element is then the linear subspace that is spanned by the frame. If one has the usual coordinate charts about $u$ in $\mathbf{R}^k$ and $x$ in $M$, so the coordinates of a particular 1-jet $j^1 x_u$ are $(u^a, x^i, x^i_a)$, then the $k$-frame at $x$ that it defines is given by:

$$\mathbf{e}_a(x) = x^i_a \frac{\partial}{\partial x^i}. \tag{2.5}$$

One can also define $m$ covectors at $u$ by way of:

$$\theta^i(u) = x^i_a \, du^a, \tag{2.6}$$

although this set of covectors does not have to be linearly independent, except when $k = m$ and one is considering 1-jets of local diffeomorphisms of $\mathbf{R}^m$ into $M$; i.e., local coordinate systems and local $m$-frames.

Dually, the manifold $J^1(M, \mathbf{R}^k)$, when restricted to submersions, for which $k \leq m$ and the rank of the differential map is always $k$, consists of *k-coframes* on $M$. If the local coordinate about a 1-jet $j^1 \phi_x$ are $(x^i, \phi^a, \phi^a_i)$ then the $k$-coframe that $j^1 \phi_x$ defines at $x \in M$ is given by:

$$\theta^a(x) = \phi^a_i \, dx^i. \tag{2.7}$$

Similarly, one can define $m$ vectors at $\phi \in \mathbf{R}^k$ by way of:

$$\mathbf{e}_i(\phi) = \phi^a_i \frac{\partial}{\partial \phi^a}, \tag{2.8}$$

but they do not have to be linearly independent unless $m = k$ and one restricts to local diffeomorphisms.

The field theories of physics generally consider 1-jets of sections of vector bundles $\pi: E \to M$, where $M$ is the spatial or space-time manifold and $E$ has fibers that carry a representation of some gauge group $G$. Furthermore, this vector bundle usually comes about by applying the "associated bundle" construction to a $G$-principal bundle $P \to M$ that one calls the *gauge structure* of the field theory and the chosen representation $G \to GL(V)$ in some vector space $V$ that becomes the model for the fibers of $E$. A $C^1$ section is then a continuously differentiable map $s: M \to E$, such that $\pi(s(x)) = x$ for every $x \in M$. One generally denotes the manifold of 1-jets of $C^1$ sections of $E \to M$ by $J^1 E$, in order to distinguish it from the 1-jets of all $C^1$ maps from $M$ to $E$ that do not have to respect the fibration.

For more general fiber bundles $B \to M$, such as $G$-principal bundles, global sections do not have to exist, and, in fact, for any principal fibration a global section exists iff it is trivial, and thus is equivalent to the projection $M \times G \to M$. However, since jets are purely local objects, and local sections of fiber bundles always exist, by the demand that they be locally trivial, one can still define $J^1B$ to be the manifold of 1-jets of local sections of $B \to M$ without addressing the existence of global sections. In particular, in the variational formulation of general relativity, one considers 1-jets of local frame fields on spacetime to be fundamental objects, so it is reassuring that the calculus of variations can still be applied, even when the spacetime manifold $M$ is not parallelizable; i.e., it does not admit a global frame field. In fact, the variation of frames – or "vierbeins" – is at the basis for the "Palatini method" of deriving Einstein's equations of gravitation from a variational principle.

### 2.2 Integrability of sections of jet projections

Each of the three basic projections that are associated with $J^1(M, N)$ admit sections, but the sections of the source projection play a special role since they are most closely associated with the issue of integrability. Namely, a general section $s : M \to J^1(M, N)$, $u \mapsto s(u)$ will have the local coordinate form:

$$s(u) = (u^a, x^i(u), x^i_a(u)). \tag{2.9}$$

If one is given a $C^1$ map $x: M \to N$, $u \mapsto x(u)$ then one can define a section $j^1x: M \to J^1(M, N)$, $u \mapsto j^1x(u)$ that one calls the *1-jet prolongation* of $x$ by essentially differentiating $x$. Thus, the coordinate form of $j^1x$ is:

$$j^1x(u) = (u^a, x^i(u), x^i_{,a}(u)), \tag{2.10}$$

in which the comma implies partial differentiation with respect to $u^a$.

There are thus sections that do not take this form, since there are $C^1$ functions from $U \subset M$ to $\mathbf{R}^{mn}$ that do not have to be partial derivatives of functions from $U$ to $\mathbf{R}^n$. Basically, one can compare the local 1-forms $\theta^i = x^i_a du^a$ to the 1-forms $dx^i = x^i_{,a} du^a$. One immediately sees that the latter are exact – hence, closed – so $d\theta^i = 0$ would be a necessary condition on the $\theta^i$ in order for them to be integrable, in this sense.

More precisely, we call a section $s : M \to J^1(M, N)$ *integrable* iff it is the 1-jet prolongation of some function $f : M \to N$:

$$s = j^1f. \tag{2.11}$$

There are two ways of characterizing the integrability of $s$: the method of contact 1-forms and the method of the Spencer operator. We shall discuss both, as well as how one relates to the other.

The manifold $J^1(M, N)$ admits $n$ canonical 1-forms $\Theta^i$, $i = 1, \ldots, n$ that we call the *contact 1-forms*. Although they can be defined globally, they are easiest to describe locally using coordinates. Basically, they take the form:

$$\Theta^i = dx^i - x^i_a du^a. \tag{2.12}$$

When one pulls them down to 1-forms on $M$ by way of any section $s$, this has the effect of replacing $x^i$ with $x^i(u)$ and $x^i_a$ with $x^i_a(u)$, so one has:

$$s^*\Theta^i = (x^i_{,a}(u) - x^i_a(u))du^a, \qquad (2.13)$$

and one sees that $s$ is integrable iff:

$$s^*\Theta^i = 0 \qquad \text{for all } i. \qquad (2.14)$$

The exterior derivatives of the contact forms are:

$$d\Theta^i = -dx^i_a \wedge du^a, \qquad (2.15)$$

and when one pulls them down to $M$ by way of $s$ the result is:

$$s^*d\Theta^i = d(s^*\Theta^i) = \tfrac{1}{2}(x^i_{a,b} - x^i_{b,a})du^a \wedge du^b. \qquad (2.16)$$

If one rewrites $\Theta^i$ as $dx^i - \theta^i$, with $\theta^i = x^i_a du^a$ then $d\Theta^i = -d\theta^i$, and the vanishing of $s^*d\Theta^i$ says that either:

$$x^i_{a,b} = x^i_{b,a} \qquad (2.17)$$

or that $d\theta^i$ vanishes. This amounts to the symmetry of mixed partial derivatives in various forms. That is, $x^i_{a,b}$ is symmetric in its lower indices iff its anti-symmetric part vanishes.

Note that since the operator $d$ commutes with that of pull-back, one must have:

$$s^*d\Theta^i = d(s^*\Theta^i), \qquad (2.18)$$

Thus, if $s^*\Theta^i$ vanishes then so does $s^*d\Theta^i$.

The *Spencer operator* is traditionally [9] defined to be:

$$D: J^1(M, N) \to \Lambda^1(M) \otimes J^0(M, N), \qquad s \mapsto s - j^1(\beta \cdot s),$$

which then needs much clarification:

1. By $J^1(M, N)$, we implicitly mean the set of $C^1$ sections of the source projection of $J^1(M, N)$.

2. By $J^0(M, N)$, we really mean vector fields that are tangent to sections of $J^0(M, N) \to M$. Locally, they take the form:

$$X = X^i(u)\frac{\partial}{\partial x^i}. \qquad (2.19)$$

Thus, the elements of $\Lambda^1(M) \otimes J^0(M, N)$ are 1-forms on $M$ with values in vector fields on the image of $x: M \to N$. They then define linear maps from each $T_u$ to the corresponding $T_{x(u)}N$, and thus contact elements to the points $x(u)$ in $N$.

3. The difference $s \mapsto s - j^1(\beta \cdot s)$ is essentially a shorthand notation for something that is easier to describe in coordinates, namely:

$$Ds(u) = (x^i_a(u) - x^i_{,a}(u))du^a \otimes \frac{\partial}{\partial x^i}. \tag{2.20}$$

Hence, the relationship between the contact 1-forms $\Theta^i$ and the Spencer operator $D$ is ultimately quite simple:

$$Ds = -s^*\Theta^i \otimes \frac{\partial}{\partial x^i}. \tag{2.21}$$

One sees that the vanishing of $Ds$ is then a necessary and sufficient condition for the integrability of $s$, as is the vanishing of $s^*\Theta^i$.

One can also investigate the consequences of the vanishing of $s^*\Theta^i$ when the section $s$ is a section of the target or contact projections, but we shall not have immediate use for the results, so abbreviate that part of the discussion.

## 3  A general class of models for physical systems

If we return to the realm of first-order contact and 1-jets then we have essentially enough mathematical machinery to make the five basic constructions that we alluded to in the Introduction to this work. We shall first present them in their abstract form and then devote the next three sections of the paper to showing how they apply in various physical contexts.

### 3.1  The five basic components of a physical model

We begin by defining a *physical system* to be a $C^1$ map $f: M \to N$ from some $m$-dimensional differentiable manifold $M$ to an $n$-dimensional differentiable manifold $N$.

1. The *kinematical state* of a physical system is a section $s: M \to J^1(M, N)$ of the source projection of $J^1(M, N)$. Thus, it locally takes the coordinate form:

$$s(u) = (u^a, x^i(u), x^i_a(u)). \tag{3.1}$$

Frequently, the kinematical state is subject to constraints, which can be holonomic or non-holonomic, perfect or imperfect. In addition to imposing boundaries on the region in $N$ that is accessible to the motion of objects, one might also define a set of equations on $J^1(M, N)$, that could have the local form:

$$C_\rho(u^a, x^i, x^i_a) = c_\rho, \qquad \rho = 1, \ldots, r, \tag{3.2}$$

in which the $c_\rho$ are constants.

One must be aware of the fact that such a set of $r$ *algebraic* equations on $J^1(M, N)$ also potentially defines a set of $r$ first-order partial *differential* equations for the functions $x^i$, when one evaluates the functions $C_\rho$ on $j^1 x$.

2. The *integrability condition* on the kinematical state of the physical system takes either the form:

$$s^* \Theta^i = \alpha^i, \tag{3.3}$$

where $\alpha^i$, $i = 1, \ldots, n$ are $n$ specified 1-forms on $M$, or:

$$Ds = -\alpha^i \otimes \frac{\partial}{\partial x^i}. \tag{3.4}$$

Most commonly the $\alpha^i$ all vanish, but the other possibility includes the essence of rotational mechanics and anholonomic frame fields, so we do not exclude it axiomatically.

The integrability of a kinematical carries with it the integrability of the constraints (3.2) to which it is subject. Basically, a constraint on the kinematical state is holonomic iff it is integrable, in the sense that the position is restricted to some submanifold of $M$. This is always the case when the functions $C_\rho$ take the form $C_\rho(u^a, x^i)$. When the constraint is non-holonomic, there are configurations $x$ such that $j^1 x$ is not constant with the given constraints, and it is possible for all points of $M$ to be accessible by motions, even though the infinitesimal generators of the motions are constrained to a proper subspace of the fiber of the contact projection.

3. A *dynamical state* of the physical system is the pull-down $s^* \phi$ of a 1-form $\phi$ on $J^1(M, N)$ by a kinematical state $s$. Thus, $\phi$ will have the local coordinate form:

$$\phi = F_i \, dx^i + \Pi^a_i dx^i_a. \tag{3.5}$$

In this expression, the $F_i$ are the components of the *generalized force* that acts on the system, and the $\Pi^a_i$ represent the components of the *generalized stress-momentum* of the system. The reason for omitting the term in $du^a$ has more to do with the calculus of variations, so we shall not elaborate at the moment, but refer to the authors previous papers [**1-5**].

The pull-down to $M$ of $\phi$ by a section $s$ of the source projection of $J^1(M, N)$ then takes the form:

$$s^* \phi = \left( x^i_{,a} F_i + x^i_{b,a} \Pi^b_i \right) du^a. \tag{3.6}$$

The duality that is associated with this definition of a dynamical state relates to the infinitesimal changes to the kinematical, which we shall discuss shortly.

4. The *constitutive law* that associates any kinematical state – or really any point of $J^1(M, N)$ itself – with a dynamical state is contained in the functional form of the components of $\phi$:

$$F_i = F_i(u^a, x^i, x^i_a), \qquad \Pi^a_i = \Pi^a_i(u^a, x^i, x^i_a). \tag{3.7}$$

5. The *balance principle* for the dynamical states of the physical system takes the form:

$$D^*\phi = 0, \tag{3.8}$$

in which the adjoint $D^*$ of the Spencer operator needs further explanation:

We regard a vector field $\delta\xi$ on the image of a section $s: M \to J^1(M, N)$ as a *virtual displacement* of the kinematical state $s(u)$; one could also regard $\delta\xi$ as a *variation* of $s$ in the usual sense. $\delta\xi$ will have the coordinate form:

$$\delta\xi(u) = \delta x^i(s(u))\frac{\partial}{\partial x^i} + \delta x^i_a(s(u))\frac{\partial}{\partial x^i_a}. \tag{3.9}$$

When one evaluates the 1-form $\phi$ on a virtual displacement $\delta\xi$, the result is a function on $J^1(M, N)$:

$$\phi[\delta\xi] = F_i\,\delta x^i + \Pi^a_i\,\delta x^i_a, \tag{3.10}$$

that can be interpreted as an infinitesimal increment of *virtual work* that is done by the virtual displacement $\delta\xi$ of the kinematical state. Thus, the 1-form $\phi$ can be thought of as dual to the virtual displacement $\delta\xi$ of a kinematical state under the canonical bilinear pairing of linear functionals and vectors that evaluates the linear functional on the vector.

Such a vector field $\delta\xi$ on $s$ is called *integrable* iff it is the 1-jet prolongation $\delta^1 x$ of a vector field on $N$ along $x : M \to N$:

$$\delta x(u) = \delta x^i(u)\frac{\partial}{\partial x^i}. \tag{3.11}$$

The *1-jet prolongation* of $\delta x$ then takes the local form:

$$\delta^1 x(u) = \delta x^i(u)\frac{\partial}{\partial x^i} + \frac{\partial(\delta x^i)}{\partial u^a}(u)\frac{\partial}{\partial x^i_a}. \tag{3.12}$$

Thus, we have, in effect, defined a linear map $j^1: \mathfrak{X}(N) \to \mathfrak{X}(J^1)$, $\delta x \mapsto \delta^1 x$.

We can call the difference between the vector fields $\delta\xi$ and $\delta^1 x$:

$$D\xi = \delta\xi - \delta^1(\beta \cdot \xi) = (\delta x^i_a - \delta x^i_{,a})\frac{\partial}{\partial x^i_a}, \qquad (3.13)$$

so $\delta\xi$ is integrable iff $D\xi = 0$, and one can always regard any $\delta\xi$ as:

$$\delta\xi = \delta^1(\beta \cdot \xi) + D\xi. \qquad (3.14)$$

The linear operator $D$ then takes vector fields in $\mathfrak{X}(J^1)$ to vector fields in $V^1_0(J^1)$, which are the vector fields on $J^1(M, N)$ that project to zero under the contact projection $\pi^1_0$. Such vector fields are then tangent to the fibers of that projection.

When $\delta\xi$ is integrable, one can go further:

$$\phi[\delta^1 x] = F_i\,\delta x^i + \Pi^a_i \frac{\partial(\delta x^i)}{\partial u^a} = \left(F_i - \frac{\partial \Pi^a_i}{\partial u^a}\right)\delta x^i + \frac{\partial(\Pi^a_i \delta x^i)}{\partial u^a}. \qquad (3.15)$$

We then introduce the notation:

$$D^*\phi = \left(F_i - \frac{\partial \Pi^a_i}{\partial u^a}\right)dx^i, \qquad (3.16)$$

which then becomes a 1-form on $N$, and we find that:

$$\phi[\delta^1 x] = D^*\phi\,[\delta x] + \frac{\partial(\Pi^a_i \delta x^i)}{\partial u^a}. \qquad (3.17)$$

Although we have denoted the operator $D^*$ as if it were adjoint to the operator $D$, more precisely, it is adjoint to $j^1$. Thus, $D^*: \Lambda^1(J^1) \to \Lambda^1(N)$.

If one defines the *total virtual work* done by the virtual displacement $\delta x$ of $s$ as:

$$W[\delta x] = \int_M \phi[\delta^1 x]V, \qquad (3.18)$$

in which $V$ is the volume element on $M$ (which we assume to be orientable), then we can say:

$$W[\delta x] = \int_M D^*\phi[\delta x]V + \int_{\partial M} (\Pi^a_i \delta x^i)\#\partial_a, \qquad (3.19)$$

in which we have introduced the notation:

$$\#\partial_a = i_{\partial/\partial u^a}V = du^1 \wedge \ldots \wedge \widehat{du^a} \wedge \ldots \wedge du^m, \qquad (3.20)$$

where the caret implies that the term under it has been omitted from the exterior product.

If the balance principle is rooted in the vanishing of the total virtual work $W[\delta x]$ dome by a virtual displacement $\delta x$ of the kinematical state $s$ that is transverse to the generalized stress-momentum (i.e., $\Pi_i^a \delta x^i = 0$) on the boundary of $M$ then the equations that result from this take the form:

$$F_i = \frac{\partial \Pi_i^a}{\partial u^a}. \tag{3.21}$$

If one uses a non-integrable virtual displacement $\delta\xi$ of $s$ then, from (3.14), one can correct (3.17) to read:

$$\phi[\delta^1 x] = D^*\phi[\delta x] + \phi[D\xi] + \frac{\partial(\Pi_i^a \delta x^i)}{\partial u^a}, \tag{3.22}$$

in which the extra term is:

$$\phi[D\xi] = \Pi_i^a(\delta x_{,a}^i - \delta x_{,a}^i) = -\Pi_i^a D\xi_a^i. \tag{3.23}$$

Suppose the $D\xi$ term is specified, *a priori*, to be a linear function of $\delta x^i$:

$$D\xi_a^i = \omega_{ja}^i \delta x^j = \omega_a^i(\delta x), \tag{3.24}$$

in which we have defined the 1-form on $N$ with values in the vector fields on $J^1(M, N)$ that are vertical – i.e., project to 0 – under the contact projection:

$$\omega_a^i = \omega_{ja}^i dx^j, \tag{3.25}$$

which corresponds to making:

$$\delta x_{,a}^i = \nabla_a \delta x^i \equiv \partial_a \delta x^i + \omega_{ja}^i \delta x^j. \tag{3.26}$$

For such a displacement, one can then regard $\phi[D\xi]$ as a correction to the partial derivatives of $\Pi_i^a$ in $D^*\phi$:

$$\phi[\delta\xi] = \bar{D}^*\phi[\delta x] + \frac{\partial(\Pi_i^a \delta x^i)}{\partial u^a}, \tag{3.27}$$

in which we have defined:

$$\bar{D}^*\phi = \left(F_i - \nabla_a \Pi_i^a\right) dx^i, \tag{3.28}$$

with:

$$\nabla_a \Pi_i^a = \frac{\partial \Pi_i^a}{\partial u^a} - \omega_{ia}^j \Pi_j^a. \tag{3.29}$$

This type (3.26) of non-integrable displacement of a kinematical state defines essentially a generalized linear connection, at least in the way that such things are introduced into the geometry of jet manifolds (see, e.g., Saunders [**8**]).

One must now alter the condition on dynamical states that defines the balance principle to take the form:

$$\bar{D}^* \phi = 0, \tag{3.30}$$

instead of (3.8).

### 3.2 Specialization to the least-action principle

Much of what we were doing above in the context of virtual work was strongly evocative of analogous procedures that one resorts to in the calculus of variations when one starts with an action functional. In fact, that is no coincidence, since the only thing that one needs to connect the two theories is to make:

$$\phi = d\mathcal{L} = \frac{\partial \mathcal{L}}{\partial x^i} dx^i + \frac{\partial \mathcal{L}}{\partial x_a^i} dx_a^i \tag{3.31}$$

in which the $C^1$ function $\mathcal{L} = \mathcal{L}(x^i, x_a^i)$ on $J^1(M, N)$ plays the role of a *Lagrangian density* for the system.

This means that we can identify:

$$F_i = \frac{\partial \mathcal{L}}{\partial x^i}, \qquad \Pi_i^a = \frac{\partial \mathcal{L}}{\partial x_a^i}. \tag{3.32}$$

One then sees that the evaluation of $\phi$ on $\delta\xi$ gives the integrand of the first variation functional

$$\delta\mathcal{L}[\delta\xi] = \phi[\delta\xi] = \frac{\partial \mathcal{L}}{\partial x^i} \delta x^i + \frac{\partial \mathcal{L}}{\partial x_a^i} \delta x_a^i, \tag{3.33}$$

and when $\delta\xi = \delta^1 x$ is integrable, one finds that:

$$\phi[\delta^1 x] = \frac{\delta \mathcal{L}}{\delta x^i} \delta x^i + \frac{\partial(\Pi_i^a \delta x^i)}{\partial u^a}, \tag{3.34}$$

in which:

$$\frac{\delta \mathcal{L}}{\delta x^i} = \frac{\partial \mathcal{L}}{\partial x^i} - \frac{\partial}{\partial u^a} \frac{\partial \mathcal{L}}{\partial x_a^i} \tag{3.35}$$

is the usual variational derivative.

This allows us to identify:

$$D^*\phi = \frac{\delta \mathcal{L}}{\delta x^i} dx^i . \tag{3.36}$$

The total virtual work that we have described then takes the form of the first variation functional for $\mathcal{L}$ when it is applied to the variation $dx$:

$$W[\delta x] = \delta \mathcal{L}[\delta x] = \int_M \left(\frac{\delta \mathcal{L}}{\delta x^i} \delta x^i\right) V + \int_{\partial M} (\Pi_i^a \delta x^i) \# \partial_a . \tag{3.37}$$

Its vanishing is a necessary, but not sufficient, condition for the action functional that is defined by $\mathcal{L}$, namely:

$$S[x] = \int_M \mathcal{L}(j^1 x) V , \tag{3.38}$$

to be a minimum for any variation $\delta x$ of $x$ that satisfies the vanishing or transverse boundary conditions on $\partial M$.

The essential difference between our starting point, which is d'Alembert's principle of virtual work[1], and Hamilton's least-action principle is an "infinite-dimensional analogue" of the difference between defining equilibrium to be a state for which the forces that act on a system vanish and defining it to be a state of minimum potential energy. That is, there are forces such as non-conservative forces and forces of non-holonomic constraint that do not admit potential energy functions, even though they still contribute to the virtual work that is done by a virtual displacement. Thus, the present variational formalism can still be applied in such cases where action functionals (or Hamiltonians, for that matter) are inappropriate.

## 4      Applications to particular physical systems

Although clearly each topic in this section could easily expand into a chapter, if not a book in its own right, we shall attempt to abbreviate that tendency by simply identifying the way that the five basic elements of the model that were introduced in the previous section. Some of the ideas presented have their roots in work done by Gallisot [**11, 12**] on the application of differential forms and jets to the mechanics of points and continua.

---

[1] A good traditional reference on d'Alembert's and Hamilton's principles is Lanczos [**10**].

### 4.1    Point mechanics

Since a point mass is assumed to be described by a time-parameterized (or proper-time-parameterized) differentiable curve in space (or spacetime), it is natural to identify the kinematical state of a moving point by a section of the source projection of $J^1(\mathbf{R}, M)$ onto $\mathbf{R}$. Such a section $s: \mathbf{R} \to J^1(\mathbf{R}, M)$ then takes the form:

$$s(t) = (t, x^i(t), v^i(t)). \tag{4.1}$$

It is integrable iff:

$$v^i = \frac{dx^i}{dt}, \tag{4.2}$$

which is also expressed by the equivalent conditions:

$$Ds = 0, \qquad s^*\Theta^i = 0, \tag{4.3}$$

since, in this case:

$$\Theta^i = dx^i - v^i\, dt. \tag{4.4}$$

The non-integrable case enters into consideration when one uses a time-rotating orthonormal frame field along the curve $x^i(t)$, instead of the natural frame field defined by the choice of coordinate system. If the angular velocity of such a moving frame is expressed by the anti-symmetric matrix $\omega^i_j(x)$ then the velocity might take the form:

$$v^i = \frac{dx^i}{dt} + \omega^i_j x^j, \tag{4.5}$$

instead of (4.2).

A non-integrable displacement of the kinematical state that corresponds to this would then take the form:

$$\delta\xi(t) = \delta x^i(t)\frac{\partial}{\partial x^i} - \omega^i_j(t)\delta x^j(t)\frac{\partial}{\partial v^i}; \tag{4.6}$$

i.e.:

$$\delta v^i = -\omega^i_j \delta x^j. \tag{4.7}$$

The dynamical state that is associated with $s(t)$ is based in a choice of 1-form in $J^1(\mathbf{R}, M)$:

$$\phi = F_i\, dx^i + p_i\, dv^i, \tag{4.8}$$

in which the components have the functional form:

$$F_i = F_i(t, x^i, v^i), \qquad p_i = p_i(t, x^i, v^i). \tag{4.9}$$

Of course, there is a physical difference between forces that are applied externally to a moving mass and forces that only come about as a result of its motion, such as viscous damping. However, one can still regard viscous damping as mathematically represented by a coupling of force to velocity, independently of any choice of curve. As for the form of the constitutive law that couples velocity to momentum, the most common choice is to choose a Euclidian metric on the spatial manifold, consider it in an orthonormal frame, which is usually the natural frame field, under the assumption of the flatness of the metric, and set the linear momentum equal to:

$$p_i(t, v^i) = m(t)\, \delta_{ij}\, v^j. \tag{4.10}$$

The possibility of mass varying with position is usually more appropriate to continuum mechanics, so we will introduce it later, but the possibility of time-varying mass is commonplace, as one uses it as the basis for jet propulsion. However, it is easy to make momentum position-dependent by considering a more interesting spatial metric $g_{ij}(x)$, in place of $\delta_{ij}$, such as when the natural frame is not orthonormal, which implies that the Levi-Civita connection of the metric is not flat. One might then set:

$$p_i(t, v^i) = m(t)\, g_{ij}(x)\, v^j. \tag{4.11}$$

When one examines the form of the adjoint integrability condition that $D^*\phi$ must vanish on the image of $s$, one finds that it is:

$$F_i(t, x^j(t), v^j(t)) = \frac{dp_i}{dt}, \tag{4.12}$$

which is, of course, Newton's second law, or the law of balance of linear momentum.

When $s$ is not integrable, one finds that the corresponding statement about $\phi$ is that the forces must include "fictitious" (i.e., frame-dependent) forces and the time derivative must be a covariant derivative:

$$\nabla_t p_i = \frac{dp_i}{dt} - \omega_i^j\, p_j. \tag{4.13}$$

The adjoint to the integrability condition on kinematical states must be altered accordingly to the vanishing of:

$$\bar{D}^*\phi = (F_i - \nabla_t p_i)\, dx^i. \tag{4.14}$$

### 4.2  Rigid body mechanics

The main difference between a point mass and a rigid body can be distilled down to the difference between a moving point and a moving frame. In fact, in its simplest manifestation, for which space is Euclidian and three-dimensional, the configuration manifold *M* can be taken to be the underlying manifold of the Lie group *ISO*(3), which is the group of rigid motions in Euclidian 3-space, or also the inhomogeneous special orthogonal group in three dimensions.

The group *ISO*(3) is the semi-direct product $\mathbf{R}^3 \times_s SO(3)$ of the three-dimensional translation group and the three-dimensional rotation group. As a manifold, one can also think of it as the (trivial) bundle $SO(\mathbf{R}^3) \to \mathbf{R}^3$ of orthonormal frames over $\mathbf{R}^3$, which then takes the form $\mathbf{R}^3 \times SO(3)$ topologically.

The actual association of a configuration of a rigid body in space with an element of *ISO*(3) generally involves making a choice of initial reference frame $(x^i(0), \mathbf{e}_i(0))$ that is fixed in the body and centered at the point $x^i(0) \in \mathbf{R}^3$, and then assuming that the time evolution of the frame takes the form:

$$(x^i(t), \mathbf{e}_i(t)) = (a^i(t), R^i_j(t))\,(x^i(0), \mathbf{e}_i(0)). \tag{4.15}$$

Here, the element $(a^i(t), R^i_j(t))$ belongs to *ISO*(3) and the action of *ISO*(3) on $SO(\mathbf{R}^3)$ takes the form:

$$(a^i, R^i_j)\,(x^i, \mathbf{e}_i) = (a^i + R^i_j x^j, R^j_i \mathbf{e}_j). \tag{4.16}$$

If one associates the initial frame with the identity element $(0, \delta^i_j) \in ISO(3)$ then the time evolution of the inhomogeneous orthonormal 3-frame $(x^i(0), \mathbf{e}_i(0))$ becomes simply a curve in *ISO*(3), which we assume to be $C^1$. Thus, the kinematical states of the rigid body can be described by sections $s(t)$ of the source projection of the jet manifold $J^1(\mathbf{R}, ISO(3))$, which then have the local coordinate form:

$$s(t) = (t, x^i(t), R^i_j(t), v^i(t), v^i_j(t)). \tag{4.17}$$

The section $s(t)$ is integrable iff:

$$v^i(t) = \frac{dx^i}{dt}, \qquad v^i_j(t) = \frac{dR^i_j}{dt}. \tag{4.18}$$

This is what the linear and angular velocity of a moving rigid body look like from an "inertial" – i.e., holonomic – frame field, such as the natural frame field of a coordinate system. They basically represent the velocity vector field $(v^i(t), \omega^i_j(t))$ that is tangent to the differentiable curve $(x^i(t), R^i_j(t))$ in *ISO*(3), which is obtained by direct time differentiation.

However, from the co-moving viewpoint of the moving frame itself $(x^i(t), \mathbf{e}_i(t))$, which is then not moving with respect to that frame, the curve degenerates to the identity element in *ISO*(3), while the velocity vectors to the curve $(x^i(t), R^i_j(t))$ must be *right-* translated back to the identity, and thus define a curve ($v_0^i(t)$, $\omega_j^i(t)$) in the Lie algebra $\mathfrak{iso}(3)$, where:

$$v_0^i(t) = \frac{dx^j}{dt}(t)\,\tilde{R}_j^i(t), \quad \omega_j^i(t) = \frac{dR_k^i}{dt}(t)\tilde{R}_j^k(t), \tag{4.19}$$

this time. Here, we are using the tilde to signify that we are using the inverse of the matrix in question, and the reason that the matrix $\tilde{R}_j^i$ is right-multiplied is due to the fact that the matrix $R_j^i$ left-multiples the initial frame $R_j^i(0)$. Such a section of $J^1(\mathbf{R}, ISO(3))$ then takes the form:

$$s(t) = (t, x^i(0), R_j^i(0), v_0^i(t), \omega_j^i(t)). \tag{4.20}$$

Thus, the concept of a non-integrable section appears quite naturally and seems to be at the basis for rotational mechanics. One then finds that when the section $s(t)$ takes the form (4.17), the effect of the Spencer operator is to give:

$$Ds = (\dot{x}^i - v^i) \otimes \frac{\partial}{\partial x^i} + \left(\dot{R}_j^i - v_j^i\right) \otimes \frac{\partial}{\partial R_j^i}. \tag{4.21}$$

This vanishes iff:

$$v^i = \dot{x}^i, \qquad v_j^i = \dot{R}_j^i. \tag{4.22}$$

Thus, in an inertial frame the integrability is equivalent to simply the idea that the velocities are, in fact, time derivatives of the position coordinates, both translational and rotational.

In the non-inertial frame, the section takes the form (4.20), so one has, upon using (4.19):

$$\dot{x}^i - v_0^i = \dot{x}^j(\delta_j^i - \tilde{R}_j^i), \qquad \dot{R}_j^i - v_j^i = \dot{R}_j^k(\delta_k^i - \tilde{R}_k^i). \tag{4.23}$$

These expressions vanish iff either the translational and rotational velocities vanish in an inertial frame, so $R_j^i(t)$ would be a constant matrix, or the curve $\tilde{R}_j^i(t)$ itself collapses to the identity, which is simply a specialization of the former condition to a particular constant matrix. Thus, the integrability of the section $s(t)$ that describes the motion in the non-inertial frame would be equivalent to the assumption that the rigid body would not be rotating, as seen from the inertial frame. But this simply says that the rotational motion is what introduces the non-integrability, which is equivalent to the non-inertial nature of the co-moving frame. In other words $\omega_j^i(t)$ is not generally composed of the time derivatives of the rotational coordinates.

The dynamical state that is associated with a kinematical state of a rigid body takes the local form:

$$\phi = F_i dx^i + \tau_i^j dR_j^i + p_i dv^i + L_i^j dv_j^i \tag{4.24}$$

in the inertial frame and:

$$\phi = \overline{F}_i dx_0^i + \overline{\tau}_i^j dR_j^i(0) + \overline{p}_i dv_0^i + \overline{L}_i^j d\omega_j^i \tag{4.25}$$

in the co-moving one.

Note that generally the empirical data that one uses to define constitutive laws are measured in the rest frame of the object. The main differences from conventional mechanical models in the present constitutive laws that are defined by the functional form of the components are:

1. Conceivably, the forces and linear momenta might pick up contributions from the angular motion, especially in the non-integrable case.

2. The generalized forces and momenta now include torques $\tau_i^j$ and angular momenta $L_i^j$, along with the linear analogues. However, the transposition of the indices relates to the fact that these matrices belong to the dual space $\mathfrak{iso}(3)^*$ of the Lie algebra $\mathfrak{iso}(3)$ and thus define linear functionals on the velocities [1].

3. Just as the mass of the rigid body might be changing in time, so might the moment of inertia, which appears when one assumes that the constitutive law that couples angular momentum to angular velocity takes the form:

$$L_i^j(t, \omega_l^k) = I_{ik}^{jl}(t) \, \omega_l^k . \tag{4.26}$$

Thus, moment of inertia represents a (potentially time-varying) linear isomorphism of the vectors spaces that underlie $\mathfrak{iso}(3)$ and $\mathfrak{iso}(3)^*$.

The effect of evaluating an element of $\mathfrak{iso}(3)^*$ on an element of $\mathfrak{iso}(3)$ is to produce an increment of instantaneous power that is being absorbed or dissipated by the rigid body in its motion. For instance, one might consider a rigid body moving in a viscous fluid within this mathematical framework. Thus, the duality principle that relates to the present model is bilinear pairing of torsors with velocities to produce virtual power, and when they are multiplied the 1-form $dt$ that parameterizes a curve, they give infinitesimal increments of virtual work.

Balance principles that follow from the vanishing of $D^*\phi$ are:

$$F_i = \frac{dp_i}{dt}, \qquad \tau_i^j = \frac{dL_i^j}{dt} \tag{4.27}$$

while the vanishing of $\overline{D}^*\phi$ gives:

---

[1] Some authors refer to elements of $\mathfrak{iso}(3)^*$ as *torsors*, which is a term that goes back to the Nineteenth Century investigations of the role of projective geometry in the mechanics of rigid bodies. While the French referred to torsors, the Germans used the term *dyname*, and the English called it a *wrench*, while the elements of $\mathfrak{iso}(3)$ were then *screws*.

$$\bar{F}_i = \nabla_t \bar{p}_i, \qquad \bar{\tau}_i^{\ j} = \nabla_i \bar{L}_i^{\ j}. \tag{4.28}$$

These then amount to the balance of linear and angular momentum in the inertial and co-moving frames, respectively.

### 4.3   Continuum mechanics

The general formalism that we introduced in Section 3 is close to what one needs in order to characterize the motion and equilibrium states of continuously-extended matter distributions. First, one needs to specify the nature of the source manifold $M$.

The role of the source manifold for extended matter is to parameterize it as a deformable region of space, somewhat like the latitudes and longitudes on a globe that is allowed to deform. Thus, the manifold $M$ represents a topological prototype, which we now denote by $\mathcal{O}$, for the various states of deformation of the object, which is embedded in a target manifold that we now call $M$. An object then takes the form of an embedding $x: \mathcal{O} \to M$, $u \mapsto x(u)$. Since one is usually expecting to integrate things defined on the object $x$, it is often convenient to further specify that the manifold $\mathcal{O}$ is a compact orientable subset of $\mathbf{R}^p$ that is expressible as a finite formal sum of singular $p$-simplexes, whether triangular or cubic. Thus, points of $\mathcal{O}$ will already consist of ordered $p$-tuples of coordinates.

An advantage of this way of describing objects in $M$ is that the parameter space $\mathcal{O}$ might also contain the time parameter, along with the ones that describe the points in space that the object occupies at a given time point. Thus a two-dimensional parameter space might describe a moving string or filament, as well as a static membrane or plate.

There is a usually a significant difference between the way that one treats solid and fluid media in continuum mechanics. In particular, fluids are usually thought of as something that is confined to channels or containers whose shape does not change in time. Thus, the story of fluid mechanics usually starts with a discussion of the flow velocity vector field, not the deformations of the medium. However, one can still apply the present formalism to the motion of individual fluid cells that get convected by the flow.

The reason for choosing embeddings is that traditionally continuum mechanics is more concerned with diffeomorphisms of regions of $M$ that, for us, will be the images of objects. If we use two embeddings $x, x': \mathcal{O} \to M$ then we can invert the first one $x$ on its image $x(\mathcal{O})$ in $M$ and then compose it with the second one $x'$ to obtain a diffeomorphism onto $f = (x' \cdot x^{-1}) : x(\mathcal{O}) \to M$ which then represents a finite deformation of the region $x(\mathcal{O})$ in the more customary sense.

The concept of a single embedding $x: \mathcal{O} \to M$ already includes the possibility of describing motion, if one assumes that $\mathcal{O}$ takes the form $[t_0, t_1] \times \mathcal{O}_s$, in which $\mathcal{O}_s$ is the spatial part of the object. The restrictions $x_0$ and $x_1$ of $x$ to the faces $t_0 \times \mathcal{O}_s$ and $t_1 \times \mathcal{O}_s$ can then be regarded as an initial and final configuration of a spatial object that is undergoing some continuous-time process. If one assumes that these restrictions are also embeddings then the composition $(x_1 \cdot x_0^{-1}) : x_0(\mathcal{O}_s) \to x_1(\mathcal{O}_s)$ is a diffeomorphism.

When $M = \mathbf{R}^m$, one can further associate a diffeomorphism $f: x(\mathcal{O}) \to M$ with a *displacement vector field:*

$$\mathbf{u}(x) = f(x) - x, \tag{4.29}$$

which is, of course, defined only locally for a more general manifold. For two embeddings $x, x'$, it takes the form:

$$\mathbf{u}(x(a)) = x'(a) - x(a). \tag{4.30}$$

Note that not every vector field on $\mathbf{R}^m$ defines the displacement vector field of a diffeomorphism. In particular, some vector fields, such as radius vector fields, would displace more than one point to the same point.

When one considers the kinematical state:

$$j^1 x(u) = (u^a, x^i(u), x^i_{,a}(u)) \tag{4.31}$$

of an object $x: \mathcal{O} \to M$, one sees that the partial derivatives involve not only generalized velocities that are associated with a time parameter – if there is one – but also spatial derivatives that describe the change of shape, as well. However, these are not directly associated with the components of the usual infinitesimal strain tensor that gets associated with the deformation $f$. Indeed, one sees immediately that the deformation $f$ is only meaningful when one is given *two* embeddings; i.e., two *states of deformation*. Thus, it represents a "relative" quantity, in the sense of "binary."

If the finite deformation[1] $f$ is described locally in coordinates by a system of equations of the form $y^i = y^i(x^j) = x^i + u^i(x)$ then the *infinitesimal strain tensor* $e_{ij}(x)$ is obtained by polarizing the matrix $u^i_{,j}(x)$ of the displacement gradient, after one lowers the upper index on $u^i(x)$ using the Euclidian metric $\delta_{ij}$ on $\mathbf{R}^m$:

$$2u_{i,j} = e_{ij} + \theta_{ij}, \tag{4.32}$$

in which:

$$e_{ij} = u_{i,j} + u_{j,i}, \qquad \theta_{ij} = u_{i,j} - u_{j,i}. \tag{4.33}$$

The anti-symmetric part $\theta_{ij}$ of $u_{i,j}$ is called the *infinitesimal rotation* of the deformation.

When the finite deformation $f$ is the result of a differentiable one-parameter family of diffeomorphisms whose infinitesimal generator is the vector field $\mathbf{v}$, one can characterize the infinitesimal strain tensor geometrically by the *Lie equation:*

---

[1] Some good references on conventional strain and stress are [**13-17**]. The last one devotes particular attention to the case of Cosserat media.

$$e_{ij} = L_v g_{ij}, \tag{4.34}$$

where the operator $L_v$ is the Lie derivative with respect to **v**, which differentiates along the integral curves of **v**, and $g_{ij}$ represents the components of the metric more generally.

Thus, the infinitesimal strain tensor measures the extent to which **v** differs from a Killing vector field, for which $e_{ij}$ would, by definition, vanish, and for which the local flow would consist of isometries of the metric $g_{ij}$. One can then regard such motions as "rigid" in the present context.

In order to relate the deformation gradient to the partial derivatives $x^i_a$, it helps to simplify the problem by making the object $x$ have the same dimension as the embedding space, so $p = m$. Thus, one can deal directly with the coordinates in all cases, and not have to introduce adapted coordinate systems for the images of the objects in space.

Another simplification is to consider only a manifold $\mathcal{O}$ that is a subset of $M$, so the initial state becomes simply the inclusion map $\mathcal{O} \to M$, $x \mapsto x$. The composition $f = x' \cdot x^{-1}$ then becomes a diffeomorphism $y: \mathcal{O} \to M$, $x \mapsto y(x)$ onto its image, and can thus be expressed locally by a system of equations in the form:

$$y^i = y^i(x^j). \tag{4.35}$$

This makes the differential matrix $dy|_x : T_x\mathcal{O} \to T_{y(x)}M$ take the local form $y^i_{,j}$.

From here, one can proceed as above to define the displacement vector field, displacement gradient, and infinitesimal strain and rotation tensors. Note that this condition easily generalizes to the non-integrable case by simply removing the commas.

An interesting aspect of the current description of extended matter is that it shows that before one deals with the usual integrability (i.e., compatibility) of the infinitesimal strain tensor, one must first address the integrability of the kinematical states of the *object*, in the sense that we are using. That is, one needs to make physical sense of what the non-integrable states of deformation would have to represent.

A section $s$ of the source projection of $J^1(\mathcal{O}, M)$ take the local form:

$$s(u) = (u^a, x^i(u), x^i_a(u)), \tag{4.36}$$

and it is integrable iff:

$$x^i_a = x^i_{,a}. \tag{4.37}$$

Now, as we pointed out above, $x^i_a(u)$ defines a set of local 1-forms $\theta^i = x^i_a(u)\,du^a$, $i = 1, \ldots, m$ on O, while $x^i_{,a}(u)$ defines the set $dx^i = x^i_{,a}(u)\,du^a$. Although they do not have to be linearly independent when $m > p$, nonetheless, one can choose an adapted coordinate system on $x(\mathcal{O})$ for which the extra 1-forms beyond $p$ vanish and one is left with two $p$-coframe fields $\theta^a$ and $dx^a$ on O. The kinematical state that is defined by $s$ is then integrable iff $\theta^a = dx^a$, so $\theta^a$ must be exact, and a non-integrable kinematical state is then associated with an anholonomic frame field $\theta^a$ on $\mathcal{O}$.

The usual way of treating the integrability of kinematical states, in the form of the infinitesimal strain tensor $e = e_{ij}\, dx^i\, dx^j$ is to say that it is *integrable* iff there is a displacement vector field **u**, or really, its metric-dual covector field $u$, such that:

$$e = 2d_s\, u, \tag{4.38}$$

where $d_s: T^*M \to S^2 M$ is the symmetrized differential operator that takes $u_i$ to $\frac{1}{2}(u_{i,j} + u_{j,i})$, locally.

The image of the first-order linear differential operator $d_s$ is not all of $S^2 M$, so not all choices of $e$ will be integrable. The integrability condition that then defines this image is given by the *Saint-Venant compatibility equations:*

$$0 = e_{ij,\,k,\,l} + e_{jk,\,l,\,i} + e_{li,\,j,\,k}. \tag{4.39}$$

These equations amount to a linearization of the condition that the deformed metric must still have vanishing Riemannian curvature, since the undeformed one did. Thus, the operator whose kernel is the image of $d_s$ is a *second*-order linear differential operator.

The dynamical state that one associates with the kinematical state is the 1-form:

$$\phi = f_i\, dx^i + \Pi_i^a dx_a^i. \tag{4.40}$$

The evaluation $\phi(\delta\xi)$ of $\phi$ on a virtual displacement $\delta\xi$ then produces the increment of virtual work:

$$\delta W = f_i\, \delta x^i + \Pi_i^a \delta x_a^i, \tag{4.41}$$

and the balance principle that is associated with the $\Pi_i^a$ takes the form:

$$f_i = \frac{\partial \Pi_i^a}{\partial u^a}. \tag{4.42}$$

Although this looks vaguely similar to the conventional balance principles of continuum mechanics, it is not identical. Thus, we should exhibit the way that this elementary set of equations relate to the more customary ones. A first step is to make the simplifications discussed above, so one can replace the $a$ index with $i, j, k$, etc., and let $u^0 = x^0 = t$ represent the time coordinate.

In the Lagrangian picture that is appropriate to solid mechanics, these balance equations take the form:

$$f_i = \frac{\partial p_i}{\partial t} + \sigma_{i,j}^j, \qquad 0 = \frac{\partial \rho}{\partial t} + (p^i)_{,i}. \tag{4.43}$$

In these equations, $p_i$ are the components of the momentum 1-form $p = p_i \, dx^i$, $\rho$ represents the mass density of the medium, and the $\sigma^i_j$ are the components of the Cauchy stress tensor.

In the Eulerian picture that is appropriate to fluid mechanics, the first set of equations takes the form:

$$L_{\mathbf{v}} p_i = -\sigma^j_{i,j} + f_i, \tag{4.44}$$

in which:

$$L_{\mathbf{v}} p_i = \frac{\partial p_i}{\partial t} + v^j \frac{\partial p_i}{\partial x^j}. \tag{4.45}$$

One can combine the two equations in (4.43) into a single equation by combining the stress, mass density, and momentum density into a single second-order tensor field $\Pi^\mu_\nu$, while we embed $f_i$ in the 1-forms on the space of one higher dimension:

$$\Pi^\mu_\nu = \begin{bmatrix} \rho & \rho v_j \\ \rho v^i & \sigma^i_j \end{bmatrix}, \quad f_\mu = \begin{bmatrix} 0 \\ f_i \end{bmatrix}. \tag{4.46}$$

Equations (4.43) then take the form:

$$f_\mu = \Pi^\mu_{\nu,\mu}, \tag{4.47}$$

which is now of the same form as (4.42).

Notice that the stress-mass-momentum tensor $\Pi^\mu_\nu$ can be regarded as dual (under the bilinear pairing of virtual work) to a generalized velocity-strain tensor of the form:

$$e^\mu_\nu = \begin{bmatrix} 1 & v_j \\ v^i & e^i_j \end{bmatrix}, \tag{4.48}$$

that is obtained by symmetrizing the generalized velocity:

$$x^\mu_\nu = \begin{bmatrix} 1 & 0 \\ v^i & x^i_j \end{bmatrix}. \tag{4.49}$$

This essentially amounts to a $C^1$ function on $\mathcal{O}$ with values in a projective-geometric representation of the group of rigid motions in $n$ dimensions.

Although the stress tensor field $\sigma = \sigma_{ij} \, dx^i \otimes dx^j$ is usually assumed to be symmetric, that assumption is based on the absence of couple-stresses in the medium, while the contrary case is that of the Cosserat medium [17, 18, 19], for which one must consider asymmetrical stress tensors. Such an asymmetry is also considered in the context of Weyssenhoff fluids [**20**], which are the classical limit of Dirac wave functions, which have non-vanishing spin and "transverse momentum." For Cosserat media, the first set of equations in (4.43) gets replaced with the two sets:

$$f_i = \frac{\partial p_i}{\partial t} + \sigma_{i,j}^j, \qquad \tau_j^i = \frac{\partial L_j^i}{\partial t} + \mu_{j,k}^{ik} + \sigma_{\cdot j}^i - \sigma_j^{\cdot i}, \qquad (4.50)$$

in which $\sigma_j^i$ is generally asymmetric and $\mu_j^{ik}$ is the couple-stress tensor.

### 4.4 Wave mechanics [**3**]

Whether one is dealing with mechanical or electromagnetic waves, one must deal with the issue of defining how one is supposed to represent a wave mathematically. For the sake of linear waves, it is usually sufficient to deal with plane waves as the building block for all other possible wave envelopes, by way of a Fourier series or Fourier transform, but when dealing with nonlinear wave motion, such a decomposition is less useful. Basically, one can start by defining a *wave* to be a "wave-like" solution to the field equations that one is addressing, and then clarify what the term "wave-like" means.

A common way of defining wave-like solutions is to assume that they take the form:

$$\psi(t, x^i) = A(t, x^i)\, e^{-i\theta(t,\,x)}, \qquad (4.51)$$

in which $A(t, x^i)$ represents the amplitude envelope of the wave and $\theta(t, x^i)$ represents its phase function. These functions can then be specialized to the more conventional applications, such as assuming that the amplitude is constant in time, or even time and space, or that the phase function is linear in the angular frequency $\omega$ and wave number $k_i$. More generally:

$$d\theta = \omega\, dt + k_i\, dx^i \equiv k. \qquad (4.52)$$

One can even generalize to the case in which the frequency-wavenumber 1-form $k$ is not exact.

As a result of a representation such as (4.51), one then obtains partial differential equations for amplitude and phase separately. The equation for $\theta$ usually takes the form of a dispersion law combined with an integrability constraint:

$$\mathrm{P}(x^\mu)[k] = \mathrm{P}_0, \qquad k = d\theta, \qquad (4.53)$$

in which $\mathrm{P}(x^\mu)[k]$ is generally a homogeneous function of $k$ of some even degree.

The most elementary dispersion law is the one associated with the linear wave equation, by way of its principal symbol, namely:

$$\eta^{\mu\nu} k_\mu k_\nu = 0 \qquad (\omega = \pm c\kappa). \qquad (4.54)$$

However, this is typical of waves that carry no mass, while the massive waves have a non-zero right-hand side to (4.54) that one might denote by $k_0^2$, which implies that there is some characteristic scale of length, such as the Compton wavelength for the electron.

The first-order partial differential equation in this case:

$$\eta^{\mu\nu} \frac{\partial \theta}{\partial x^\mu} \frac{\partial \theta}{\partial x^\nu} = 0 \tag{4.55}$$

is the homogeneous four-dimensional form of the "eikonal" equation of geometrical optics.

One can already sense that one is dealing with sections of the source projection of $J^1(M, \mathbf{R})$ if one calls its local coordinates $(x^\mu, \psi, \psi_\mu)$. However, when the wave function decomposes into the product form (4.51), it would be more appropriate to use $J^1(M, \mathbf{R}^2)$ with the local coordinates $(x^\mu, A, \theta, A_\mu, k_\mu)$. More generally, one might wish to make the amplitude function $A$ take the form of a section of some complex vector bundle over $M$ that one might regard as a "bundle of local oscillators," so $e^{-i\theta(t, x)}$ is a complex scalar function that multiplies its elements. Here, we shall regard a kinematical state of a wave as a section $s(x)$ of the source projection of $J^1(M, \mathbf{R}^2)$, which then has the coordinate form:

$$s(x) = (x^\mu, A(x), \theta(x), A_\mu(x), k_\mu(x)). \tag{4.56}$$

An integrable section of that source projection will then satisfy:

$$A_\mu = A_{,\mu}, \qquad k_\mu = \theta_{,\mu}. \tag{4.57}$$

The dispersion law then takes the form of a function $P$ on $J^1(M, \mathbf{R}^2)$ that is independent of $\theta$, which then defines the first-order partial differential equation above by way of $P(j^1\theta) = P_0$. For linear dispersion laws, it will also be independent of $A$ and $A_\mu$. Thus, if one regards a section of the source projection of $J^1(M, \mathbf{R})$ as a kinematical state of the wave then the dispersion law amounts to a constraint on the kinematical state.

One sees that the kinematical state of a wave, which is essentially due to $k$, is dual to the kinematical state of a point particle, which is essentially due to $\mathbf{v}$, under the bilinear pairing $k(\mathbf{v})$, which has the dimension of frequency. Here, the points in question will be the points of instantaneous wave fronts, for which both $t$ and $\theta$ are constant.

One can associate a spatial group velocity $\mathbf{v}_g$ with $k$ by way of the dispersion law:

$$v_g^i = -\frac{\partial P / \partial k_i}{\partial P / \partial \omega}, \tag{4.58}$$

although the "space" that this refers to is the projective space $\mathbf{RP}^3$, not the affine space $\mathbf{R}^3$, since the form of the $v_g^i$ is (minus) that of the inhomogeneous coordinates of $\mathbf{RP}^3$ that are associated with the homogeneous coordinates $(P_\omega, P_i)$, with the obvious definitions of $P_\omega$ and $P_i$ that come from the components of $dP$.

A dynamical state for the wave would then take the form of a 1-form $\phi$ on $J^1(M, \mathbf{R}^2)$ that does not involve $dx^i$. It then takes the form:

$$\phi = \rho_A \, dA + \rho_\theta \, d\theta + p_A^\mu dA_\mu + p_\theta^\mu dk_\mu. \tag{4.59}$$

in which the functions $\rho_A$, $\rho_\theta$ serve as sources for the amplitude and phase waves, while $p_A^\mu$ and $p_\theta^\mu$ define the components of their generalized momenta, which are now vector fields, instead of 1-forms. Of course, when the amplitude takes its values in a vector space or manifold of dimension higher than one, both $\rho_A$ and $p_A^\mu$ will have more than one component index accordingly.

The functional form of the components of $\phi$:

$$\rho = \rho(x^\mu, A, \theta, A_\mu, k_\mu), \qquad p^\mu = p^\mu(x^\nu, A, \theta, A_\nu, k_\nu) \tag{4.60}$$

in which we generically skip the subscripts $A$ and $\theta$, embody the constitutive laws for the wave motion. These laws are usually derived from empirical considerations that become specific to the medium in which the waves are propagating.

An infinitesimal virtual displacement $\delta\xi$ of the kinematical state $s$ is then associated with the function on $J^1(M, \mathbf{R}^2)$:

$$\phi[\delta\xi] = \rho_A \, \delta A + \rho_\theta \, \delta\theta + p_A^\mu \delta A_\mu + p_\theta^\mu \delta k_\mu, \tag{4.61}$$

but the physical interpretation of this function is ambiguous until one specifies the nature of the amplitude.

If $\delta\xi = \delta^1\theta$ is integrable then:

$$\phi[\delta\xi] = (D^*\phi)_A \, \delta A + (D^*\phi)_\theta \, \delta\theta, \tag{4.62}$$

up to a divergence, and the balance principles for the amplitude and phase take the form:

$$\rho_A = \frac{\partial p_A^\mu}{\partial x^\mu}, \qquad \rho_\theta = \frac{\partial p_\theta^\mu}{\partial x^\mu}. \tag{4.63}$$

It might clarify the foregoing process somewhat to see how it works in some familiar cases.

To begin with, consider the effect of the d'Alembertian operator:

$$\Box\psi \equiv \frac{\partial^2\psi}{\partial t^2} - c^2 \delta^{ij} \frac{\partial^2\psi}{\partial x^i \partial x^j}, \tag{4.64}$$

on wave functions of the form (4.51). By direct computation, one finds that:

$$\Box\psi = [1/A \, \Box A - \eta^{\mu\nu} k_\mu k_\nu - i(2/A \, \eta^{\mu\nu} A_\mu k_\nu + \Box\theta \,]\psi. \tag{4.65}$$

Naturally, this equation must be defined only on the support of the wave, where $A$ is non-vanishing. We are also assuming that $A$ is real, although the complex case does allow one to deal with the dissipation of amplitude in absorptive media.

This latter equation takes the form of a generalized eigenvalue problem for which the eigenvalue $\lambda$ is a function on $J(M, \mathbf{R}^2)$. If one lets $\lambda$ take the form:

$$\lambda = \lambda_A - k_0^2 - i(\alpha_0^2 + \rho_\theta), \tag{4.66}$$

in which $k_0^2$ and $\alpha_0^2$ are real constants, then one derives the following systems of equations for the amplitude and phase:

$$\Box A = \lambda_A A, \quad 2\eta^{\mu\nu} A_\mu k_\nu = \alpha_0^2 A, \quad \Box\theta = \rho_\theta, \quad \eta^{\mu\nu} k_\mu k_\nu = k_0^2. \tag{4.67}$$

These equations then take the form of two differential equations for $A$ and $\theta$ and two constraints on the section $s(x)$, as it is described in (4.56). Furthermore, one can replace the d'Alembertian operators with divergences:

$$\Box A = \frac{\partial}{\partial x^\mu}\left(\eta^{\mu\nu} \frac{\partial A}{\partial x^\mu}\right) = \frac{\partial p_A^\mu}{\partial x^\mu}, \quad \Box\theta = \frac{\partial p_\theta^\mu}{\partial x^\mu}. \tag{4.68}$$

When $\lambda_A = \alpha_0^2 = \rho_\theta = 0$, equations (4.67) become two partial differential equations and two algebraic equations for $A_\mu$ and $k_\mu$ alone. However, if one regards $A_\mu$ as $A_{,\mu}$ and $k_\mu$ as $\theta_{,\mu}$ then all of the equations are partial differential equations for $A$ and $\theta$.

One can construct a dynamical state for the wave $\psi$ that gives the differential equations of (4.67) as balance principles in the form of:

$$\phi = (\lambda_A A)\, dA + \rho_\theta\, d\theta + p_A^\mu\, dA_\mu + p_\theta^\mu\, dk_\mu. \tag{4.69}$$

In order to put wave equations into the preferred form (4.63), it helps to think of them as being presented in their "conservation law" form [**21**], rather than their second-order form. One can not only specialize the dynamical state (4.69) to give numerous special cases of common interest in wave theory, but when one takes a closer look at $\eta^{\mu\nu}$ one sees that it really plays the role of a constitutive law for the medium, which then becomes linear, isotropic, and homogeneous, under the assumption that $c$ is a constant, and non-dispersive. In the general case, any of these assumptions are dubious, and one sees how the alternative possibilities, such anisotropy or nonlinearity, all follow from more detailed assumptions about the functional form of (4.60), or $\eta^{\mu\nu}$, for that matter.

It is intriguing that the usual de Broglie relations $p_\mu = \hbar k_\mu$ takes the form of a constitutive law if one ignores the fact that the $p_\mu$ are still the components of a 1-form, not a vector field, such as one might get from $p^\mu = \hbar g^{\mu\nu} k_\nu$.

### 4.5    Electromagnetism

Electromagnetism, as it is commonly practiced, generally involves the integrability of a different differential operator than simple differentiation, namely, the exterior derivative operator $d: \Lambda^k M \to \Lambda^{k+1} M$. However, other formulations of electromagnetism have been proposed in which the integrability of other operators is the issue.

Most relativistic treatments of electromagnetism tend to trivialize the contribution of the electromagnetic constitutive law that associates the "kinematical" states in the form of field strength 2-forms $F$ on the four-dimensional spacetime manifold $M$, which is then given a Lorentzian metric $g$, *a priori*, with the "dynamical" states, in the form of the electromagnetic excitation bivector fields $\mathfrak{h}$. However, the "pre-metric" approach to electromagnetism [**22, 23**] focuses more attention on the nature of that map, while the Lorentzian metric becomes a consequence of the dispersion law for electromagnetic waves when one has such a constitutive law in effect. Thus, we shall use that formulation of the Maxwell equations in order to illustrate the features of the model as we have defined it above.

First, one identifies the kinematical state of an electromagnetic field as the field strength 2-form $F$ on $M$:

$$F = \theta^0 \wedge E + \#\mathbf{B}, \tag{4.70}$$

in which $\theta^0$ represents a 1-form whose annihilating hyperplanes in each tangent space define the "spatial" subspace, $E$ is the electric field strength 1-form, $\mathbf{B}$ is the magnetic field strength bivector field, and $\#: \Lambda_2 M \to \Lambda^2 M$, $\mathbf{A} \mapsto i_\mathbf{A} V$ is the Poincaré isomorphism that is defined by a choice of volume element:

$$V = dx^0 \wedge dx^1 \wedge dx^2 \wedge dx^3 = \frac{1}{4!} \varepsilon_{\kappa\lambda\mu\nu} dx^\kappa \wedge dx^\lambda \wedge dx^\mu \wedge dx^\nu \tag{4.71}$$

on $M$, which is assumed to be orientable. In this last expression, the symbol $\varepsilon_{\kappa\lambda\mu\nu}$ refers to the Levi-Civita symbol, which is equal to +1 whenever the indices $\kappa\lambda\mu\nu$ are an even permutation of 0123, − 1 whenever they are an odd permutation, and 0 otherwise.

Thus, the isomorphism takes the local form:

$$(\#\mathbf{A})_{\kappa\lambda} = \tfrac{1}{2} \varepsilon_{\kappa\lambda\mu\nu} A^{\mu\nu}. \tag{4.72}$$

The integrability condition for a kinematical state is then given by the first Maxwell equation:

$$dF = 0, \tag{4.73}$$

which is the necessary condition for $F$ to admit an electromagnetic potential 1-form $A$:

$$F = dA. \tag{4.74}$$

Two comments must then be made at this stage:

1. The condition (4.73) is sufficient only when all closed 2-forms on *M* are exact; i.e., when the two-dimensional de Rham cohomology vector space $H^2_{dR}(M)$ vanishes. By de Rham's theorem [**24, 25**], this is equivalent to saying that the real singular cohomology vector space $H^2(M; \mathbf{R})$ vanishes in two dimensions. The non-vanishing of this vector space is the basis for the possibility of magnetic monopoles.

2. The 1-form *A* is not unique, but only up to the addition of any closed 1-form. Locally, every closed 1-form is exact, so one usually sees the arbitrary closed 1-form expressed as $d\lambda$, and the smooth function then becomes a *gauge function*, which must be chosen whenever one chooses an *A*.

The dynamical states, as we just mentioned, take the form of electromagnetic excitation bivector fields:

$$\mathfrak{h} = \mathbf{e}_0 \wedge \mathbf{D} + \#^{-1} H, \tag{4.75}$$

in which the vector field $\mathbf{e}_0$ defines the "temporal" subspace of each tangent space, which is transversal to the spatial one, **D** is the electromagnetic excitation (or displacement) vector field, and *H* is the magnetic excitation 2-form.

An electromagnetic constitutive law then takes the form of a map $\chi: \Lambda^2 M \to \Lambda_2 M$ that takes the fiber of $\Lambda^2 M$ at each $x \in M$ to the corresponding fiber of $\Lambda_2 M$ at *x* in a diffeomorphic way. If one wishes to simplify somewhat then one can specify that it is also an invertible linear map, although that will rule out a lot of nonlinear optics and effective field theories for quantum electrodynamics. Indeed, one is already restricting oneself to "non-dispersive" media, for which the operator that takes 2-forms to bivector fields would be an integral operator, not an algebraic one.

In the usual metric formulation of electromagnetism, which treats only the case of the classical electromagnetic vacuum, the constitutive law takes the form of the linear isomorphism of 2-forms with bivector fields that is defined by "raising both indices" using the Lorentzian metric *g*. That is:

$$\mathfrak{h}^{\mu\nu} = \sqrt{-g}\; g^{\mu\kappa} g^{\nu\lambda} F_{\kappa\lambda}, \tag{4.76}$$

which makes:

$$\chi^{\kappa\lambda\mu\nu} = \tfrac{1}{2} \sqrt{-g}\; (g^{\mu\kappa} g^{\nu\lambda} - g^{\nu\kappa} g^{\mu\lambda}). \tag{4.77}$$

The appearance of the factor $\sqrt{-g}$ is due to the fact that when one is dealing with a Lorentzian manifold, it is customary to alter the definition of the volume element that was defined above in (4.71) so that the volume is invariant under all frame changes, and not just orthonormal ones, which then makes the new volume element take the form:

$$V_g = \sqrt{-g}\; dx^0 \wedge dx^1 \wedge dx^2 \wedge dx^3 = \frac{1}{4!} \sqrt{-g}\; \varepsilon_{\kappa\lambda\mu\nu}\, dx^\kappa \wedge dx^\lambda \wedge dx^\mu \wedge dx^\nu. \tag{4.78}$$

The duality principle between kinematical states and dynamical states is then defined simply by the evaluation of the 2-form $F$ on the bivector field $\mathfrak{h}$:

$$F(\mathfrak{h}) = \chi(F, F), \tag{4.79}$$

which then defines the electromagnetic field Lagrangian density.

The differential operator that is adjoint to $d$ under the Poincaré isomorphism # is the *divergence* operator $\delta: \Lambda_k M \to \Lambda_{k-1} M$ that is defined by:

$$\delta = \#^{-1} \cdot d \cdot \#. \tag{4.80}$$

One finds that it does, in fact, agree, with the usual divergence operator on vector fields.

The conservation law that is dual to the integrability of $F$ is given by Maxwell's second equation, in its pre-metric form:

$$\delta\mathfrak{h} = \mathbf{J}, \tag{4.81}$$

in which $\mathbf{J}$ represents the electric current vector field, whose temporal component represents the static electric charge density for the time-space splitting of the tangent bundle $T(M)$ that is defined by $(\mathbf{e}_0, \theta^0)$. However, since $\delta^2 = \#^{-1} \cdot d \cdot \# = 0$, not all vector fields $\mathbf{J}$ are acceptable as source currents, but only the ones with vanishing divergence, which then amounts to the conservation of charge.

The equations of pre-metric electromagnetism are then collectively:

$$dF = 0, \qquad \delta\mathfrak{h} = \mathbf{J}, \qquad \mathfrak{h} = \chi(F), \qquad \delta\mathbf{J} = 0. \tag{4.82}$$

Since the last equation looks like an "integrability condition" for $\mathbf{J}$, one might also start with it as the fundamental equation and then treat the second set as a sort of "dual gauge" condition, which then suggests that the bivector field $\mathfrak{h}$ is, in that picture, analogous to the potential 1-form $A$. However, the difference is that the bivector field $\mathfrak{h}$ represents physically measurable quantities, such as polarizations, while only potential differences are measurable.

## 5  Discussion

Although the basic elements of a physical model, as we defined them in the language of jet manifolds, seems to be consistently applicable to many common physical examples of broad generality, nonetheless, it is in the last two applications to wave mechanics and electromagnetism that there still seems to be some uncertainty about the most advantageous choice of application. This is especially the case in electromagnetism, since the conventional approach is more based in the integrability of differential forms under the exterior derivative operator. Similarly, the conventional compatibility equations for symmetric infinitesimal strain tensors are not as simple as the vanishing of a Spencer operator. However, one might confer Pommaret [**26**] for some idea of how one might redefine continuum mechanics in such a way that integrability does involve only the Spencer operator is its various guises.

Still, the fact that the methodology is clearly rooted in the first principles of physical law is encouraging that it might prove to be consistently useful in the cause of theoretical models.

**References**

1. D. H. Delphenich, pre-print: arXiv:1109.0461.
2. D. H. Delphenich, Ann. Phys. (Berlin) **18** (2009), 1-22.
3. D. H. Delphenich, Ann. Phys. (Berlin) **18** (2009), 206-230.
4. D. H. Delphenich, Ann. Phys. (Berlin) **18** (2009), 45-56.
5. D. H. Delphenich, pre-print: arXiv:0708.1572.
6. C. Ehresmann, Comptes rendus de l'Acad. Sci. **233** (1951), 598-600, 777-779, 1081-1083; **234** (1952), 1028-1030, 1424-1425.
7. C. Ehresmann, in *Colloque internationale de géométrie différentielle* (C. N. R. S, 1953), 97-110.
8. D. J. Saunders, *The Geometry of Jet Bundles* (Cambridge University Press, Cambridge, 1989).
9. D. C. Spencer, Bull. Amer. Math. Soc. **75** (1969), 179-239.
10, C. Lanczos, *The Variational Principles of Mechanics*, 4$^{th}$ ed. (Dover, Mineola, NY, 1970).
11. F. Gallisot, Ann. Inst. Fourier (Grenoble) **8** (1958), 291-335.
12. F. Gallisot, Ann. Inst. Fourier (Grenoble) **4** (1952), 145-297.
13. A. E. Green and W. Zerna, *Theoretical Elasticity* (Oxford University Press, Oxford, 1954).
14. L. D. Landau and E. M. Lifschitz, *Theory of Elasticity* (Pergamon, London, 1959).
15. C. Truesdell and R. Toupin, "The Classical Field Theories of Physics," in *Handbuch der Physik* III/1, ed. S. Flügge (Springer, Berlin, 1960), pp. 226-793.
16. F. D. Murnaghan, *Finite Deformation of an Elastic Solid* (Wiley, NY, 1951).
17. P. P. Constantinescu, *Dynamics of Linear Elastic Bodies* (Abacus, Tunbridge Wells, 1975).
18. E. Cosserat and F. Cosserat, *Théorie des corps déformables* (Hermann, Paris, 1909)
19. E. Kröner, ed., *Mechanics of Generalized Continua,* Proc. of 1967 IUTAM Symposium in Freudenstadt and Stuttgart (Springer, Berlin, 1968).
20. J. Weyssenhoff and A. Raabe, Acta Phys. Pol. **9** (1947), pp. 7-18, 19-25, 26-33, 34-45, 46-53.
21. A. Jeffrey and T. Taniuti, *Nonlinear Wave Propagation, with applications to physics and magnetohydrodynamics* (Academic Press, NY, 1964).
22. F. W. Hehl and Y. N. Obukhov, *Foundations of Classical Electrodynamics* (Birkhäuser, Boston, 2003).
23. D. H. Delphenich, Ann. Phys. (Leipzig) **14** (2005), 347-377.

24. G. de Rham, *Variétés différentiables* (Hermann, Paris, 1955); Eng. transl. by F. R. Smith, *Differentiable Manifolds* (Springer, Berlin, 1984).
25. R. Bott and L. Tu, *Differential Forms in Algebraic Topology* (Springer, Berlin, 1982).
26. J.-F. Pommaret, *Lie Pseudogroups in Mechanics* (Gordon and Breach, NY, 1988).